\documentclass[8pt,aps]{revtex4}
\usepackage{amssymb}
\usepackage{latexsym}

\usepackage{epsfig}

\usepackage{dcolumn}
\usepackage{amsmath}
\usepackage{amsfonts}
\usepackage{bm}
\usepackage{color}

\newcommand{\be}{\begin{equation}}
\newcommand{\ee}{\end{equation}}
\newcommand{\bea}{\begin{eqnarray}}
\newcommand{\eea}{\end{eqnarray}}

\newcommand{\p}{\partial}
\newcommand{\s}{\sigma}

\newcommand{\la}{\langle}
\newcommand{\ra}{\rangle}
\newcommand{\rd}{\mbox{d}}
\newcommand{\ri}{\mbox{i}}
\newcommand{\re}{\mbox{e}}

\renewcommand{\vec}[1]{{\bm #1}}

\begin{document}
\title{Fractionalized Fermi Liquid in a Kondo-Heisenberg  Model}
\author{ A. M. Tsvelik}
\affiliation{Condensed Matter Physics and Materials Science Division, Brookhaven National Laboratory, Upton, NY 11973-5000, USA}
 \date{\today } 
\begin{abstract}
The Kondo-Heisenberg model is used as a controllable tool to demonstrate the  existence  of a peculiar metallic state with unbroken translational symmetry where the Fermi surface volume is not controlled by the total electron density. I use a  non-perturbative approach where the strongest interactions are taken into account by means of exact solution, and corrections are controllable. In agreement with the general requirements formulated in (T. Senthil {\it et.al.}  Phys. Rev. Lett. {\bf 90}, 216403 (2003)), the resulting metallic state represents a fractionalized Fermi liquid where well defined quasiparticles coexist with gapped fractionalized collective excitations. The system undergoes a phase transition to an ordered phase (charge density wave or superconducting), at the transition temperature which is parametrically small in comparison to the quasiparticle Fermi energy. 
\end{abstract}


\maketitle

\section{Introduction}

 There is a compelling experimental evidence that in the so-called underdoped regime of  the cuprate materials the  volume of the quasiparticle Fermi surface (FS) is not proportional to the number of electrons \cite{arpes1,arpes2,arpes3,stm,harrison,louis}. 
Below a certain temperature,  an  arc-like  FS is seen only in the nodal regions and the antinodal ones are gapped (the pseudogap phenomenon). It is likely that such arcs are  part of closed Fermi pockets with unevenly distributed spectral weight \cite{yrz2}. 

  Attempts to explain this phenomenon  arising as a result of  FS reconstruction caused by  a broken  translational symmetry (see, for example, \cite{norman}) are not satisfactory. The most obvious symmetry breaking mechanism would be charge density wave (CDW) formation, however, the CDW occupies only a small part of the phase diagram where  the pseudogap is  observed \cite{hucker,blanco}. The measurements of the Hall coefficient also indicate that the pseudogap and the CDW are different phenomena \cite{louis}. The fact that the wave vectors of the CDW connect the tips of the Fermi arcs \cite{cdw2} also points to the conclusion that the CDW emerges from the pre-existing pseudogap state. These facts  point to the  existence of a peculiar metallic state with a small FS (herein, the SFS state). 


 These experimental facts pose a fundamental problem about status and validity of the Luttinger theorem. Do they point towards its violation or just to our misunderstanding of this fundamental theorem? According to  
\cite{agd,chub,essler1,igor} it is the latter since in reality Luttiger theorem (LT) relates the  total particle density not to the volume enclosed by the Fermi surface, but to the volume  enclosed by the surface in momentum space where  the single electron Green's function at zero frequency  $G(\omega =0, {\bf k})$ changes sign.  This allows   both poles (the FS) and zeroes of the Green's function to contribute to LT.  
The classic illustration of this thesis  given  in \cite{agd}, is the Green's function of a BCS superconductor where translational symmetry is unbroken:
\be
G(\omega,k) = \frac{\omega +\epsilon(k)}{\omega^2 - \epsilon(k)^2 - \Delta^2}.
\ee
The Bogolyubov quasiparticles do not have a FS; however, $G(0,k)$ changes sign exactly at $\epsilon(k)=0$ and  LT is fulfilled solely through the  zeroes of the Green's function. These considerations, together with \cite{krt}, laid the foundation for the phenomenological electron Green's function suggested for the pseudogap state of the cuprates by Rice and co-workers (the YRZ Green's function) \cite{yrz} which has turned out to be very successful in explaining various experiments (see, for instance, \cite{arpes3,yrz2,carbotte}).  

 Unfortunately, the surface of zeroes of the Green's function is not observable and hence cannot serve as a test for SFS. However, Oshikawa has given a different perspective on LT by reformulating it   in a  form 
where the electron density is related to the Friedel oscillations which may exist even without a FS \cite{oshikawa}. The validity of Oshikawa's theorem and its consistency with the traditional version of LT has been demonstrated for one-dimensional (1D) strongly correlated models \cite{essler1,zachar,cdw} including those where the FS is altogether absent. 
The electron density reveals itself in the wave vector of the CDW, as suggested in \cite{oshikawa}, as well as  in the zeroes of $G(0, \pm k_F)$\cite{cdw}. 

  For dimensions $D>1$  it  was suggested on the basis of the Oshikawa's theorem that the existence of SFS metal requires a topologically nontrivial ground state and  fractionalized excitations (for instance, electrically neutral spin S=1/2 ones) \cite{subir1}. This is a very interesting idea though  attempts to demonstrate the existence of SFS  in  microscopical models (see, for example, \cite{subir2,subir3} where it is called FL*) have relied on gauge theory approaches which involve too many uncontrollable steps. 

 In this paper I present a nonperturbative controllable solution of a strongly correlated model in $D>1$ which satisfies LT in both of the formulations outlined above. It has a topologically nontrivial ground state, fractionalized excitations and  a SFS of sharp quasiparticles, alongside  a few other  remarkable properties similar to those observed in the underdoped cuprates. The model describes an array of ladders with one leg being spin S=1/2 Heisenberg chain and the other filled by a weakly interacting Fermi gas. There is a direct tunneling between electronic chains of different ladders. The system undergoes a phase transition to an ordered phase (CDW or superconducting), but in $D>2$ the transition temperature $T_c$ is parametrically small in comparison to the quasiparticle Fermi energy $\epsilon_{F,qp}$. The latter fact allows one to formally consider the limit $T_c/\epsilon_{F,qp} \rightarrow 0$ to make sense of the discussion of LT. 



\section{The core model: the Kondo-Heisenberg (KH) chain} 
 A single  KH chain consists  of  an antiferromagnetic spin S=1/2 Heisenberg chain (HC) coupled to a one dimensional electron gas (1DEG) via an anferromagnetic exchange interaction:
\bea
H = \sum_k \epsilon(k) \psi^+_{k\s}\psi_{k\s} + \frac{J_K}{2} \sum_{k,q}\psi^+_{k+q,\alpha}\vec\s_{\alpha\beta}\psi_{k,\beta} {\bf S}_q + J_H\sum_n {\bf S}_n{\bf S}_{n+1}, \label{model1}
\eea
where $\psi^+,\psi$ are creation and annihilation operators of the 1DEG, $\s^a$ are the Pauli matrices, ${\bf S}_n$ is the spin S=1/2 operator on site $n$ and ${\bf S}_q$ is its Fourier transform.   It is assumed that $J_K << J_H$ and  the 1DEG is far from half filling, $|2k_Fa_0 -\pi| \sim 1$. Under these assumptions one can formulate the low energy description of (\ref{model1}), taking into account that the backscattering processes between excitations in the HC and the 1DEG are suppressed by the incommensurability of the 1DEG.  The effective theory is valid for energies much smaller than both the Fermi energy $\epsilon_F$ and the Heisenberg exchange interaction  $J_H$. The model (\ref{model1}) is  integrable \cite{zachar} and   I will use the exact solution  as a springboard for a controllable approach to the model  of an array of   KH chains in dimension $D>1$.

 I start with   the linearization of the spectrum of the 1DEG: $\epsilon(k) \approx \pm v_F(k \mp k_F)$ and $\psi(x) = \re^{-\ri k_F x} R(x) + \re^{\ri k_F x}L(x)$. To take advantage of the SU(2)$_{charge}\times$SU(2)$_{spin}$  symmetry of the model I will express  the low energy Hamiltonian in terms of generators of the   SU$_1$(2) Kac-Moody algebra (non-Abelian bosonization, see Appendix A and \cite{books}). These are the spin ($F^a$) and charge ($I^a$)  current operators of the 1DEG $F^a_{R}=\frac{1}{2}R^+\s^a R, F^a_L = \frac{1}{2}L^+\s^a L$ and $I^z_R = R^+_{\s}R_{\s}, ~~I^+_R= R^+_{\uparrow}R^+_{\downarrow}, ~~I^-_R = R_{\downarrow}R_{\uparrow}$ (with similar  for $I^a_L$ with $R \rightarrow L$) and ${\bf j}_R,{\bf j}_L$  of the HC.  Currents of different type and chirality commute. 
In the continuum limit we have  
\bea
{\bf S}_n = [{\bf j}_R(x) + {\bf j}_L(x)] + (-1)^n {\bf N}_s(x) +..., ~~ x= na_0\label{S}
\eea
where the dots stand for less relevant operators, $a_0$ is the lattice distance, ${\bf N}_s(x)$ is the staggered magnetization operator defined in Appendix A. Likewise, the spin density of the 1DEG is   
\bea
\frac{1}{2}\psi^+\vec\s\psi(x) =  {\bf F}_R + {\bf F}_L + \Big[\re^{2\ri k_F x}\Delta_{cdw} + H.c.\Big]+..., \label{sdensity}
\eea
where $\Delta_{cdw}$ operator is defined in Appendix A. 
Substituting (\ref{S},\ref{sdensity}) into (\ref{model1}) we find that since $2k_F \neq \pi/a_0$, the oscillatory terms in (\ref{S},\ref{sdensity}) drop out and the final result contains  only the spin currents: 
$
J_K({\bf F}_R +{\bf F}_L)({\bf j}_L + {\bf j}_R) $. 
Moreover, its the relevant part contains  only products of the currents of different chirality; the marginal interaction $
V_{marg} = J_K({\bf F}_R{\bf j}_R + {\bf F}_L{\bf j}_L)$ can be dropped as the first approximation. The resulting Hamiltonian is  

\bea
&& {\cal H}_{eff} = {\cal H}_{charge} 
+ {\cal H}_s^{(Rl)} + {\cal H}_s^{(Lr)} \label{model3}\\
&& {\cal H}_{charge} = \frac{2\pi v_F}{3}\Big( :{\bf I}_R{\bf I}_R: + :{\bf I}_L{\bf I}_L:\Big) \label{charge}\\
&& {\cal H}_s^{(Rl)} = \frac{2\pi v_F}{3}:{\bf F}_R{\bf F}_R: + \frac{2\pi v_H}{3}:{\bf j}_L{\bf j}_L: + J_K {\bf F}_R{\bf j}_L\label{GN1}\\
&& {\cal H}_s^{(Lr)} = \frac{2\pi v_F}{3}:{\bf F}_L{\bf F}_L: + \frac{2\pi v_H}{3}:{\bf j}_R{\bf j}_R: + J_K {\bf F}_L{\bf j}_R \label{GN2}
\eea
Here $v_F,v_H = \pi J_H/2$ are the Fermi velocity of the 1DEG and the spinon velocity of the HC respectively. The double dots denote  normal ordering. 
The model (\ref{model3}) was studied in \cite{zachar}  and then in \cite{kivelson}. 
Its  most remarkable features are  (i) the decoupling of the spin sector  into  two independent sectors with different parity leading to  (ii) the emergent high SU(2)$_{charge}\times$SU(2)$_{spin}\times$SU(2)$_{spin}$ symmetry.  The latter feature explains a high degeneracy of the order parameter manifold.

 The charge sector  (\ref{charge}) is critical, the spectrum is linear: $\omega = v_F|k|$. The spin sector is described by  integrable model (\ref{GN1},\ref{GN2}), at $J_K >0$ their spectrum consists of gapped spin 1/2 excitations (spinons) \cite{andrei} with dispersion relations $E(k)_{Lr} = E(-k)_{Rl} =E(k)$ (see Fig.1):
\bea
 E(k) = k(v_H-v_F)/2 +\sqrt{k^2(v_F+v_H)^2/4 + \Delta^2}, \label{disp}
\eea
where $
\Delta = C\sqrt{J_KJ_H}\exp[- \pi(v_F + v_H)/J_K], $ with $C$ being a nonuniversal numerical factor.

\begin{figure}[!htb]
\centerline{\includegraphics[ angle = 0,
width=0.5\columnwidth]{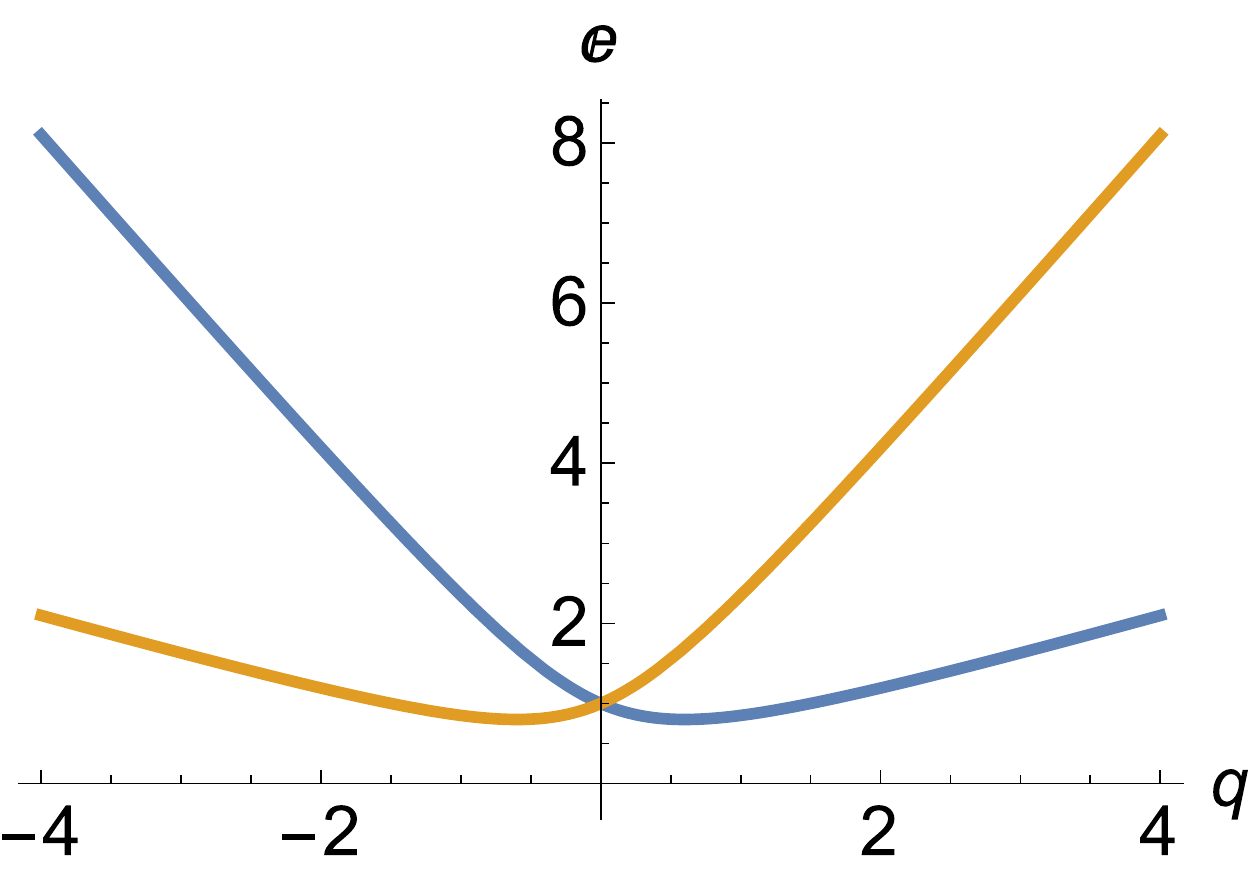}}
\vspace{-.50cm}
\caption{The dispersion of the solitons  in the KH chain (\ref{disp}). $e= E/\Delta$, $q=k_x(v_Hv_F)^{1/2}/\Delta$ and  $v_F/v_H = 1/4$.}
 \label{Spec}
\end{figure}

As is clear from (\ref{GN1},\ref{GN2}), the spinon gaps are generated by paring of spinons of a given chirality from the 1DEG with their partners of opposite chirality from the HC. Such entanglement of chiral modes makes the  ground state topologically nontrivial (Appendix B).  Since the spinons from the 1DEG do not pair to each other, there are no order parameters (OPs)  formed solely from the electronic operators \cite{remark}. Instead, there are composite OPs whose correlation functions have a power law decay:
\bea
&& {\cal O}_{cdw} = \psi^+(x)\Big[({\bf S}_x{\bf S}_{x+a_0})\hat I + \ri(\vec\s {\bf S}_x)\Big]\psi(x)\re^{\ri(\pi/a_0 +2k_F)x}\label{OP1}\\
&& {\cal O}_{sc} =  \ri(-1)^{x/a_0}\psi(x)\s^y\Big[({\bf S}_x{\bf S}_{x+a_0})\hat I + \ri(\vec\s{\bf S}_x)\Big] \psi(x),\label{OP2}\nonumber
\eea
where $\hat I$ is a unit matrix. OP (\ref{OP1})  displays Friedel oscillations with the wave vector  $Q = \pi/a_0 + 2k_F$ which  includes both localized and band electrons in agreement with \cite{oshikawa}. These OPs can be conveniently written in the matrix form:
\bea
\hat {\cal O} = \left(
\begin{array}{cc}
{\cal O}_{cdw} & {\cal O}_{sc}^+\\
-{\cal O}_{sc} & {\cal O}_{cdw}^+\\
\end{array}
\right) = 
 A\hat g, \label{g}
\eea
where $A\sim \Delta$ is an amplitude and $g$ is the matrix field of the SU$_1$(2) Wess-Zumino-Witten-Novikov model (\ref{charge}) (see Appendix A for details). 





\subsection{ Correlation functions}. 

To proceed we will need explicit expressions for some correlation functions of the KH model. They are fixed largely by  symmetry considerations,  with a minimal knowledge of the spectrum and the operator structure of the theory, as was done in \cite{essler1,cdw,SM}. 

 The corresponding calculations are more easily done with a use of Abelian bosonization as explained in Appendix A. Such bosonization is possible due to the unique property of SU$_1$(2) WZNW model, namely  due to its equivalence to the theory of free massless bosons (\ref{gauss}) and the fact that both the fermion operators (\ref{fermions}) and primary fields of the Heisenberg chain can be factorized into products of chiral operators (\ref{z}).  Such factorization does not generally occur in conformal theories. According to  (\ref{fermions},\ref{RL}) the single particle Green's function factorizes into a product of two independent functions determined by the charge and the spin sector respectively. Thus for the right movers we have:
\bea
G_{RR} = \la\la z_-^c(\tau,x)[z^c_-]^+(0,0)\ra\ra \la\la z_{\s}^s(\tau,x)[z^s_{\s}]^+(0,0)\ra\ra, \label{GRR}
\eea
with a similar expression for the left movers with $z$ substituted for $\bar z$. The latter operators are chiral and can be expressed in terms of bosonic exponents (\ref{z}).
Likewise, for the staggered magnetizations of the Heisenberg chain we have 
\bea
\la\la {\bf N}(\tau,x){\bf N}(0,0)\ra\ra = \la\la z^H(\tau,x)[z^H(0,0)]^+\ra\ra \la\la \bar z^H(\tau,x)[\bar z^H(0,0)]^+\ra\ra \label{NN}
\eea 

Since the charge sector is described by the Gaussian  noninteracting theory (\ref{gauss}), the corresponding correlator in (\ref{GRR}) is easy to calculate:
\bea
\la\la z_-^c(\tau,x)[z^c_-]^+(0,0)\ra\ra \sim (v_F\tau -\ri x)^{-1/2} \label{holon}
\eea
The next problem is to calculate the correlators of the spin components which enter in (\ref{GRR}) and (\ref{NN}). 
 As was explained in \cite{ZL,essler}, most of the spectral weight in these correlators  comes from the processes with an emission of a single massive spinon. Therefore it is sufficient to calculate just one matrix element. This was done using the Lorentz symmetry considerations \cite{essler}. Such considerations are directly applicable for the case $v_H =v_F$, but the general situation can be continuously deformed into the Lorentz invariant one. 

 I will use the relativistic parameterization of the spectrum (\ref{disp}):
\bea
&& p= [\Delta/(v_F+v_H)]\sinh\theta, \\
&& E_{Rl}(\theta) = \Delta(v_F\re^{\theta} + v_H\re^{-\theta})/(v_F+v_H), ~~ E_{rL}(\theta) = \Delta(v_H\re^{\theta} + v_F\re^{-\theta})/(v_F+v_H),\\
&& E_{lR}(\theta) = E_{Rl}(-\theta), ~~E_{Lr}(\theta) = E_{rL}(-\theta).\nonumber
\eea
where parameter $\theta$ is called rapidity. Lorentz transformation corresponds to a shift of rapidities of all particles by the same amount. Since $z_{\s},\bar z_{\s}$ have Lorentz spins 1/4 and -1/4 respectively, their matrix elements  between the vacuum and a state with a single soliton with rapidity $\theta$ are determined by the Lorentz invariance and are given by 
\bea
\la \theta,-\s[|z_{\s}(0,0)]^+|0\ra = Z_0^{1/2}\re^{\theta/4}, ~~ \la \theta,\s|[\bar z_{\s}(0,0)]^+|0\ra = Z_0^{1/2}\re^{-\theta/4},
\eea
where $Z_0$ is a nonuniversal prefactor.

The difference between the matrix elements of the Heisenberg and the 1DEG spinon operators is in their coordinate dependence. For instance, we have 
\bea
&& \la \theta,-\s|[z_{\s}^s(\tau,x)]^+|0\ra = Z_0^{1/2}\re^{\theta/4}\exp[-\tau E_{Lr}(\theta) -\ri xp(\theta)], \nonumber\\
&& \la \theta,-\s|[z_{\s}^H(\tau,x)]^+|0\ra = Z_0^{1/2}\re^{\theta/4}\exp[-\tau E_{lr}(\theta) -\ri xp(\theta)],
\eea
Substituting these matrix elements into the Lehmann expansion for the correlation functions(\ref{GRR},\ref{NN}) and using (\ref{holon}) we arrive at the following expressions.
The single electron Green's functions are similar to the ones in the Hubbard model and in the model of the 1DEG with attractive interaction  \cite{essler1, mou}.
For small $|k| << k_F$ we have $G(\omega, k\pm k_F) = G_{RR, LL}(\omega,k), ~~ G_{RR}(\omega,k) = G_{LL}(\omega,-k)$,
\bea
 && G_{RR}(\omega,k) = \label{Green}\\
&& \frac{Z_0}{\omega -v_Fk}\Big[\frac{\Delta}{\sqrt{-(\omega -v_Fk)(\omega+v_Hk) + \Delta^2}}-1\Big] +...,\nonumber
\eea
where the dots stand for terms with emission of more than one spinon and $Z_0$ is a nonuniversal numerical factor.
At  $\omega =0$ the Green's function changes sign by going through  zero at wave vectors $\pm k_F$. Since the Green's function for the localized electrons has zeroes at $\pm \pi/2a_0$,  the total volume  inside of the surface of zeroes  is $\pi/a_0 + 2k_F$ in a full agreement with LT.

 Due to the decoupling of the spin sector  (\ref{GN1},\ref{GN2}) the correlators of the staggered parts of the magnetizations are products of the spinon Green's functions. The dynamical magnetic susceptibilities have the following form \cite{SM}:
\bea
&& \la\la {\bf N}(\tau,x){\bf N(0,0)}\ra\ra = \label{chi}\nonumber\\
&& \frac{Z_N}{\pi\sqrt{v_H^2\tau^2 + x^2}}\exp\Big\{-\Delta[(\tau + \ri x/v_H)(\tau - \ri x/v_F)]^{1/2}\Big\}\times\nonumber\\
&& \exp\Big\{- \Delta[(\tau - \ri x/v_H)(\tau + \ri x/v_F)]^{1/2}\Big\}, \nonumber\\
&& \la\la {\bf s}(\tau,x){\bf s}(0,0)\ra\ra = \frac{Z_e \cos(2k_Fx)}{k_F\sqrt{v_F^2\tau^2 + x^2}}\la\la {\bf N}(\tau,x){\bf N(0,0)}\ra\ra, \label{chiDEG}
\eea
where $Z_N,Z_e$ are nonuniversal numerical factors. In the frequency -momentum space the susceptibility of the local spins (\ref{chi}) displays a strong continuum centered at $k = \pm \pi/a_0$ and the susceptibility of the 1DEG (\ref{chiDEG}) displays a weaker continuum around $\pm 2k_F$. At $v_H=v_F =v$ the Fourier transform of (\ref{chi}) is  
\be
\chi_{1D} = \frac{ Z_N}{\sqrt{4\Delta^2 + (vk_x)^2- \omega^2}} \label{chi1}
 \ee

\section{Coupling the ladders}. 

The goal of this paper is to construct a workable model in $D>1$ which would display SFS with robust quasiparticles. I consider an array of KH chains coupled together by  a direct tunneling between the 1DEGs and an exchange between the HCs. For simplicity I consider only the nearest neighbor interactions and take into account the most relevant part of the exchange interaction: 
\bea
&& H_{tunn} = t\sum_y \int \rd x (\psi^+_y(x)\psi_{y+1}(x) + H.c.),  \label{tunn}\\
&& H_{ex} = \sum_y \tilde J\int \rd x {\bf N}_y(x){\bf N}_{y+1}(x), \label{exchange}
\eea
The exchange integral $\tilde J$ is a sum of the antiferromagnetic  superexchange between the HCs and the ferromagnetic one 
   generated in the 2nd order in the interchain tunneling $\tilde J_{ferro} \sim - J_K^2t^2/W^3$.
Hence $\tilde J$  can have any sign and magnitude and hence is an independent parameter. This fact constitutes  a distinct feature of the model.

\begin{figure}
\centerline{\includegraphics[angle = 0,
width=0.5\columnwidth]{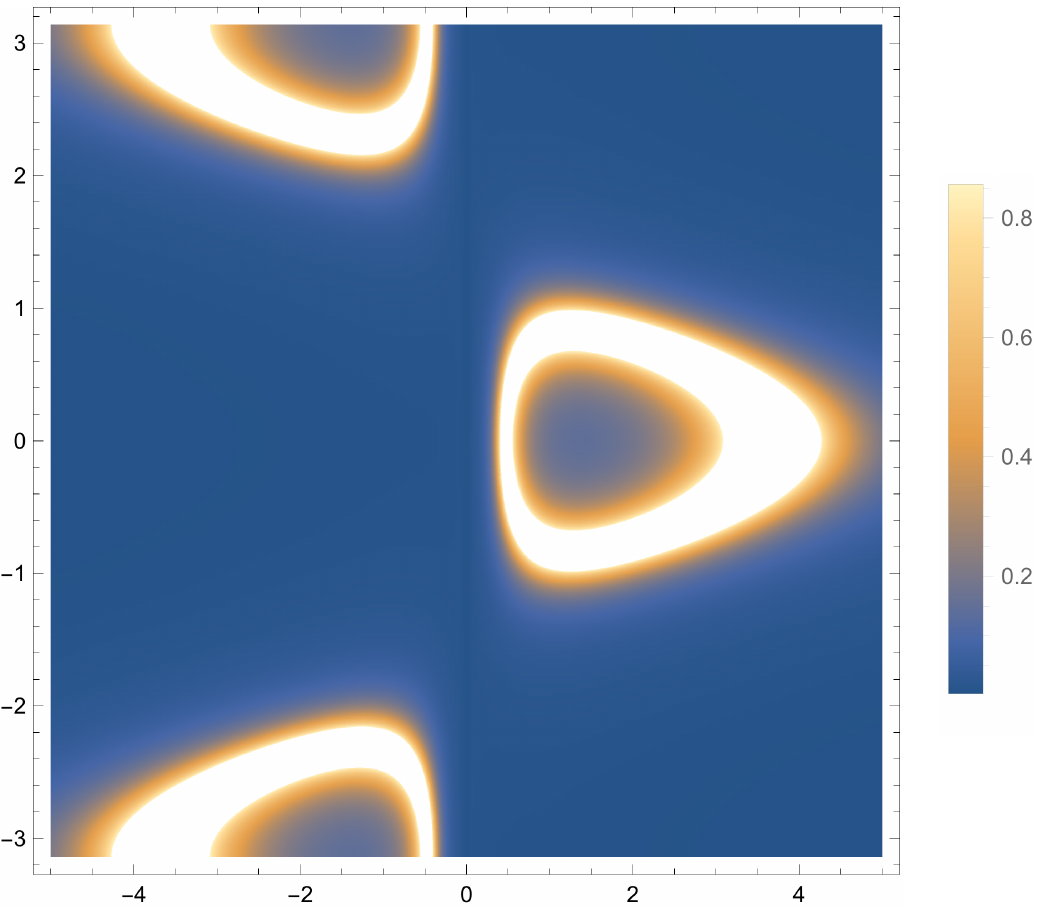}}
\caption{The plot of the quasiparticle weight near $k_x= k_F$ for $t_0(v_H/v_F)^{1/2}/\Delta = 5$ and $v_F/v_H =0.1$. The vertical axis is $k_yb$, the horizontal is $q=(k_x-k_F)(v_Hv_F)^{1/2}/\Delta$. The picture for $k_x <0$ is the mirror image of this one. }
 \label{FS3}
\end{figure}

\begin{figure}
\centerline{\includegraphics[angle = 0,
width=0.6\columnwidth]{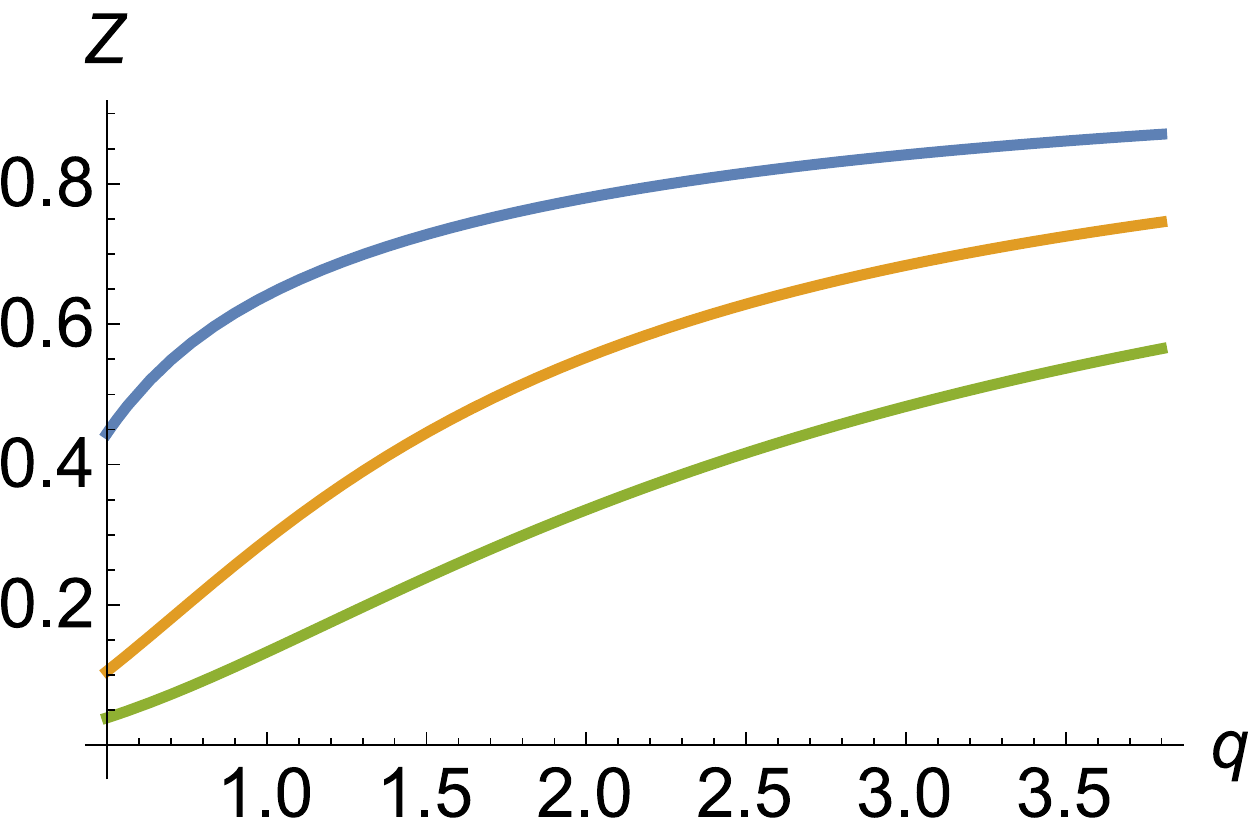}}
\caption{The quasiparticle residue Z as a function of  $q=k_x(v_Hv_F)^{1/2}/\Delta$ for (from top to bottom) $v_F/v_H = 3, 1, 0.1$.}
 \label{Z}
\end{figure}

In the Random Phase approximation (RPA) tunneling (\ref{tunn}) modifies the electron Green's function: 
\bea
G(\omega, {\bf k}) = [G_{1D}^{-1}(\omega,k_x) - t({\bf k})]^{-1}, \label{rpa}
\eea
where ${\bf k} = (k_x,k_y)$, $G_{1D}$ is given by (\ref{Green}) and $t({\bf k})$ is the Fourier transform of the interchain hopping amplitude. Since $G_{1D}(\omega)$ has a strong singularity at the threshold of the spinon-holon continuum, the interchain tunneling leads to the creation of holon-spinon bound states (quasiparticles) \cite{essler1}. The quasiparticle (QP) branch splits from the continuum and ones the tunneling amplitude exceeds the critical level $
|t(k_y)|  > 3.33 \Delta(v_F/v_H)^{1/2}$ 
electron and hole Fermi pockets appear as on Fig(\ref{FS3}) \cite{fail}. The QP FS is defined by  $\tilde t\cos (k_yb) = \pm q[1-(q^2+1)^{-1/2}]^{-1}$ with $q = k(v_Fv_H)^{1/2}/\Delta, ~~\tilde t = 2t(v_H/v_F)^{1/2}/\Delta$. The QP residue changes around the FS:
\bea
Z(q)/Z_0 = \frac{(1-(1+q^2)^{-1/2})}{\Big\{1+ \frac{q^2}{2}(\frac{v_F}{v_H}-1)(q^2+1)^{-1}[(q^2+1)^{1/2}-1]^{-1}\Big\}}, \nonumber
\eea
 For $v_F < v_H$ the change of  the QP spectral weight is significant (see Figs.(\ref{FS3},\ref{Z}) and the pockets may look as arcs. 


\subsection{Corrections to the RPA and the stability of the QPs. } 

One may wonder whether the quasiparticles  are robust against corrections to RPA. In the previous works on the coupled chain arrays \cite{essler1},\cite{essler2} we developed a formal scheme to take into account corrections to RPA. It was demonstrated that dangerous corrections originate from three kinds of physical processes. First, there are processes which generate a coupling between staggered magnetizations of different chains (\ref{exchange}). Second, there is a coupling between composite OPs generated in higher orders of perturbation theory in the interchain hopping and exchange interactions. Both types of processes can lead to ordering and  translational symmetry breaking. This in turn will cause a reconstruction of the FS. The third type of dangerous corrections is related to interactions between the quasiparticles and the gapless collective modes. 

Let us consider the first process. The SFS quasiparticles are robust against the onset of magnetic order if $|\tilde J| < \Delta$ 
so that the magnetic ordering with a subsequent FS reconstruction is prevented. As was demonstrated in the previous subsection 
in the present  model    $\tilde J/\Delta$ is an independent parameter and therefore such regime of stability can be easily accomodated. The interchain interaction is suppressed due  to the peculiar nature of the KH chain ground state, where the right- and left-moving electron degrees of freedom of the given chain are virtually decoupled and the spinon modes of the 1DEG have their pairing partners on the Heisenberg chains.
So we do not need to worry about the magnetic order. 


Now let us address the corrections to RPA related to the coupling between the composite OPs. This coupling is also parametrically small because 
   a coupling of the composite OPs necessarily includes the spin operators from the Heisenberg chains and, as soon as there is no coupling between the localized spins the OPs do not couple either. The coupling appears only if  there are interactions between the spins from different chains or there is an external field applied to one of the Heisenberg chain order parameters. This issue will be addressed in more detail in the following Section. 
   
   As I have mentioned above, one have to watch for yet another kind of potentially dangerous corrections, namely the interactions between the quasiparticles mediated by the gapless collective modes. Indeed, the Fermi surfaces of particles and holes on (\ref{FS3}) are nested and hence can be gapped. However, the coupling in the present case is weak because the wave vectors of the  COPs do not connect these Fermi surfaces. 
   
   So we can conclude that there is a temperature range below the quasiparticle chemical potential where the RPA is qualitatively robust. This temperature range can be made as wide as one wishes by varying parameters of the model. One can make corrections to RPA numerically small by making the interchain hopping long range, as it was done in (\cite{essler1,essler2}), but I consider this as unnecessary pedantry. 

 \subsection{ Collective excitations.} 

In a contrast to coupled S=1/2 HCs where the spinons undergo confinement \cite{shelton}, the fractionalized excitations of the KH spin liquid are  robust against the interchain exchange (\ref{exchange}). The spinons of KH model are gapped and the interaction (\ref{exchange})  creates their bound states  in the gap, with the spinons surviving  as the incoherent  continuum. This can be seen in  the dynamical susceptibilities:
\bea
\chi(\omega, k_x+\pi/a_0,k_y) = \{[\chi_{1D}(\omega,k_x)]^{-1} - 2\tilde J\cos k_yb\}^{-1}, \label{chi2}
\eea
where $b$ is the lattice distance in the transverse direction and the one-dimensional susceptibility $\chi_{1D}$ at $v_H=v_F =v$ is given by  (\ref{chi1}). The poles of (\ref{chi2}) give the  spectrum of the coherent spin S=1 bound states:  $
E^2= (vk_x)^2 + (2\Delta)^2 - [2\tilde J Z_N\cos( k_yb)]^2.$
 The spectrum  remains gapped  and the ground state translationally invariant as soon as  $|\tilde J| < \Delta$. Otherwise the magnetic excitations condense and a commensurate spin density wave (SDW) order emerges. Its wave vector is either $(\pi/a_0,\pi/b)$ ($\tilde J >0$) or $(\pi/a_0,0)$ ($\tilde J<0$).  

\section{Ordering and the phase diagram.} 

The main aim of this paper is to demonstrate a possibility of existence of the SFS metallic state as a degenerate Fermi liquid. In this context the problem of order is a peripheral issue and will be addressed briefly. The main point is to demonstrate that the transition temperature is an independent parameter not related to the quasiparticle Fermi energy. 

In contrast to the SDW instability the model admits  types of order which occur for any interaction strength. They are driven  by  the coupling of the critical collective charge modes through  a coupling of  OPs (\ref{OP1},\ref{OP2}). The corresponding operator is the   primary field of the  SU$_1$(2) Wess-Zumino-Novikov-Witten (WZNW) model (\ref{charge}). It is $2\times 2$ matrix $g$ which elements include the CDW and $s$-wave superconducting order parameters (\ref{g}). 
Due to their composite nature such a coupling  also requires to couple the staggered magnetizations of the HCs and  one needs  the exchange interaction  (\ref{exchange}) to facilitate it. Alternatively, one can apply an external periodic potential along the HC which couples to the dimerization operator: $V = \epsilon \sum_n (-1)^n({\bf S}_n{\bf S}_{n+1})$. Such potential is naturally present in a 3D layered  array of crisscrossed chains like the one existing  in La$_{1.875}$Ba$_{0.125}$CuO$_4$ \cite{tranquada1}. In this situation the dimerization would strengthen the coupling between the OPs within the layers, but the interlayer coupling would not be possible \cite{tranquada2} so that  the ordering will be purely 2D, like the one observed in \cite{tranquada1}.

The effective Ginzburg-Landau action for the collective modes is written in terms of the SU(2) matrix field $g$ (\ref{g}). The partition function is the path integral $Z = \int Dg \exp(-S)$   with action 
\bea
 S = \sum_y\Big[ W[g_y] - {\cal J}\int_0^{1/T} \rd\tau\int \rd x \mbox{Tr}(\s^zg_y\s^zg^+_{y+1} +H.c.)\Big], \label{GL}
\eea 
where $W[g]$ is the SU$_1$(2) WZNW action whose explicit form is given in Appendix A and the estimate  ${\cal J} \sim {\tilde J}(t/\Delta)^2$ is derived in Appendix C. In the Hamiltonian formulation $W[g]$ corresponds to (\ref{charge}). Notice that the sign of ${\cal J}$ coincides with the sign of the exchange interaction.  

  Since $\s^zg\s^z$ matrix also belongs to SU(2) group, on a bipartite lattice with nearest neighbor interactions action (\ref{GL}) possesses the global SU(2) symmetry. The  transformation $\s^z g \s^z \rightarrow g$ changes sign of the superconducting (SC) order parameter on one sublattice. Since the scaling dimension of $g$ is 1/2, the interchain coupling is strongly relevant and  the mean field transition temperature is $T_c^{MF} \sim |{\cal J}|$. This estimate also gives the maximal possible value for the zero temperature stiffness. 
  
  Due to the  non-Abelian symmetry of the action in  $D=2$  the increased fluctuations shift the transition to $T=0$. In a quasi 2D array of parallel chains the transition is shifted back to finite $T_c < T_c^{MF}$, but the degeneracy between the superconductivity and CDW   will manifest itself  as enhanced fluctuations at $T_c< T < T_c^{MF}$ \cite{Efetov_Pepin}.  $T_c^{MF}$ increases with  $\tilde J$, becoming larger with the  approach to  the magnetic Quantum Critical point. Since in the current model $\tilde J$ is an independent parameter the energy scale associated with the ordering does not compete with Fermi energy of the QPs.

  Depending on the sign of $\tilde J$ the OP manifold includes  either a $(2k_F, \pi/b)$ composite CDW and a $(\pi/a_0,0)$ composite  SC $(\tilde J <0)$ or a $(2k_F,0)$ CDW and the $(\pi/a_0,\pi/b)$  SC. The degeneracy between these OPs can be lifted by, for example, forward scattering interactions in the 1DEG which violate the particle-hole SU(2) symmetry in the charge sector.  The coupling ${\cal J}$ can be further increased by  application of an external potential which couples to the staggered energy density $\gamma (-1)^n({\bf S}_n{\bf S}_{n+1})$. Then  its the fusion of interchain tunneling operators  (\ref{tunn}) 
\be
 t [\psi^+_y(x)\psi_{y+1}(x) + H.c.] = t(R^+_yR_{y+1} + L^+_yL_{y+1} + H.c.)
\ee 
contributes $\delta {\cal J} \sim t^2\gamma^2/\Delta^3$.

\section{Conclusions and Acknowledgements}. 

In accordance with the main aim of this paper the results for the KH array demonstrate  a possibility of a SFS state. They also conform to the earlier theoretical ideas \cite{subir1}:  the model has a topologically nontrivial ground state and   gapped fractionalized spin  excitations.  As an additional bonus the results display  many features associated with   the underdoped cuprates giving rise to a hope that  the underlying physics is the same. Indeed, the normal state of the model  has a SFS with an unevenly distributed spectral weight and the volume not related to the electron density. The phase diagram contains  superconducting and CDW orders and a magnetic QCP. The ordering is preceded by strong fluctuations and since it is not driven by quasiparticles, one should expect a small stiffness. Moreover, the stiffness is unrelated to the particle density as was recently emphasized in the experiments \cite{bozovic}. Due to the staggered nature of the OPs, the coupling between OPs in perpendicular chains is greatly weakened, which  provides a mechanism for the effective dimensional reduction observed in arrays of crisscrossed chains \cite{tranquada2}.  Ones the order  is established the Fermi pockets become gapped due to the proximity effect.  


Besides these general properties which persist throughout the entire parameter range of the model, there are those which emerge for particular parameter values. For instance, when the bare qiasiparticles are slower than the spinons $v_F << v_H$ the QP spectral weight changes significantly along the FS which may create an impression of `` Fermi arcs''. The small FS state is unstable against ordering, but since the KH chain possesses a rich set of OPs,  the nature of the order can be vary depending on secondary interactions. Thus in the presence of  a staggered scalar potential, which emerges naturally in the layered configuration of crossed KH chains (as in the striped LaBaCuO \cite{tranquada2}),  there will be a competition between CDW and SC order. 


I am grateful to A. Chubukov, N. Robinson, T. M. Rice, A.-M. Tremblay, G. Kotliar, R. M. Konik, P. D. Johnson, J. Tranquada for interesting discussions and encouragement. The work was  supported by the U.S. Department of Energy (DOE),  Division  of Condensed Matter Physics and  Materials Science, under Contract No. DE-AC02-98CH10886.

\appendix

\newpage


\section{Bosonization}

Below I describe the most important properties of the SU$_1$(2) Wess-Zumino-Novikov-Witten (WZNW) model and its relation to the S=1/2 Heisenberg chain and 1DEG.

 This model can be expressed as a model of free bosons (see (\ref{gauss}) below) and also in the form used in the main text. The latter one expresses the Hamiltonian in terms of the current operators. 
The currents of the same chirality and type  satisfy the SU$_1$(2) Kac-Moody algebra:
\bea
[j_R^a(x), j_R^b(x')] = \ri\epsilon^{abc}j^c_R(x)\delta(x-x') + \frac{\ri}{4\pi}\delta_{ab}\delta'(x-x'),
\eea
with the same commutation relations for the left currents $j^a_L$. The electron spin ${\bf F}_R = \frac{1}{2}R^+\vec\s R, ~~ {\bf F}_L = \frac{1}{2}L^+\vec\s L$ and charge currents ${\bf I}$ satisfy the same algebra.

Besides the Hamiltonian representation used in the main text, the WZNW model admits the Lagrangian representation. In particular, the SU$_1$(2) WZNW action  is given by 
\bea
W[g] = \frac{1}{16\pi}\int \rd\tau \rd x \mbox{Tr}(\p_{\mu}g^+\p_{\mu} g) - \frac{\ri}{24\pi}\int_0^{\infty}\rd\xi \int \rd\tau\rd x\epsilon^{\alpha\beta\gamma}\mbox{Tr}(g^+\p_\alpha gg^+\p_\beta gg^+\p_\gamma g).
\eea

 The advantage of using the WZNW representation is that it makes the SU(2) symmetry manifest. However, for many practical calculations is convenient to use the Abelian bosonization. This is possible since the SU$_1$ Kac-Moody algebra admits an Abelian representation. The corresponding Hamiltonian (for instance, (\ref{charge})) can be written as the Hamiltonian of free bosons:
\bea
{\cal H}_{charge} = \frac{v_F}{2}\Big[(\p_x\Theta_c)^2 +(\p_x\Phi_c)^2\Big], \label{gauss}
\eea
where  field $\Phi_c$ and its dual field $\Theta_c$ satisfy the standard commutation relations $[\Phi_c(x),\p_x\Theta_c(x')] = \ri\delta(x-x')$. Likewise the Hamiltonians for the spin sector of 1DEG and the S=1/2 Heisenberg chain can be written in the same Gaussian form with bosonic fields $\Phi_s,\Theta_s$ and $\Phi_H,\Theta_H$ respectively. The SU(2) symmetry imposed on the Gaussian model manifests itself in the selection of the operators constituting the operator basis of the theory (see below).  

 Model (\ref{gauss}) is critical, the excitation spectrum is linear. Hence its Hilbert space factorizes into holomorphic and antiholomorphic parts. This agrees with the fact that the WZNW Hamiltonian can be written as a sum of commuting parts containing currents of different chirality. In fact, the Gaussian model (\ref{gauss}) has a unique property among the critical models: its primary fields can be factorized into a product of holomorphic and antiholomorphic parts containing exponents of holomorphic $\varphi_a$ and antiholomorphic $\bar\varphi_a$ ($a= c,s, H$ parts of the bosonic fields 
\be
\varphi = (\Phi+\Theta)/2, ~~ \bar\varphi =   (\Phi-\Theta)/2.
\ee
For instance,  the bosonization rules for the fermion operators are 
\bea
R_{\s} = \frac{\xi_{\s}}{\sqrt{2\pi a_0}}\re^{-\ri\sqrt{2\pi}(\varphi_c + \s\varphi_s)}, ~~ L_{\s} = \frac{\xi_{\s}}{\sqrt{2\pi a_0}}\re^{\ri\sqrt{2\pi}(\bar\varphi_c + \s\bar\varphi_s)}, \label{fermions}
\eea
where $\xi_{\s}$ are Klein factors $\{\xi_{\s},\xi_{\s'}\} = 2\delta_{\s\s'}$. 

 As I have mentioned above, the SU(2) symmetry manifests itself in the selection of the operators. The operator basis contains only derivatives of fields $\varphi, \bar\varphi$ and  integer powers of the  exponents 
\bea
&& z_{\s} = (2\pi a_0)^{-1/4}\exp[\ri\s\sqrt{2\pi}\varphi], ~~ \bar z_{\s} = (2\pi a_0)^{-1/4} \exp[-\ri\s\sqrt{2\pi}\bar\varphi], ~~ \s = \pm 1.\nonumber\\
&& z_{\s} = z^+_{-\s}. \label{z}
\eea
The chiral fields $z_{\s}^a, \bar z_{\s}^a$ have conformal dimensions (1/4,0), (0,1/4) respectively and can be considered as the holon and the spinon operators of the 1DEG ($a= c,s$)  and the spinon operators  of the Heisenberg chain ($a=H$). According to (\ref{fermions}) the annihilation operators of the right- and left moving electrons can be written as 
\bea
R_{\s} = \xi_{\s}\Big(z^c_{-}z^s_{\s}\Big), ~~ L_{\s} = \xi_{\s}\Big(\bar z^c_{-}\bar z^s_{\s}\Big). \label{RL}
\eea
The operator $\Delta_{cdw} = R^+_{\s}L_{\s} = (\bar z^c_+ z^c_-)(\bar z^s_{-\s}z^s_{\s})$. 

 In the  S=1/2 Heisenberg model also possesses an approximate symmetry between correlation functions of the staggered components of the energy density and the magnetization operators such that they can be united in a single SU(2) matrix field
\bea
\hat G(x) = (-1)^n\Big[a({\bf S}_n{\bf S}_{n+1}) + \ri b(\vec\s{\bf S}_n)\Big], ~~ x= a_0n,
\eea
where $a,b$ are nonuniversal amplitudes. The symmetry is not perfect due to the marginally irrelevant current-current interaction. This field is the spin 1/2 primary field of the SU$_1$(2) WZNW model. It can be factorized:
\bea
G_{\s\s'} = \frac{1}{\sqrt 2} \re^{\ri\pi(1-\s\s')/4}z^H_{\s}[\bar{z^H}_{\s'}]^+. \label{G}
\eea

Models (\ref{GN1},\ref{GN2}) have their own OPs with nonzero vacuum expectation values. 
\bea
\la {\cal O}_{rL}\ra = \sum_{\s}\la z^s_{\s}[\bar{z^H}_{\s}]^+\ra, ~~ \la {\cal O}_{lR}\ra = \sum_{\s}\la [{\bar z^s}_{\s}]^+ z^H_{\s}\ra. \label{ops}
\eea
They  form the amplitude of the composite OPs (\ref{g}):
\be
A = \la {\cal O}_{rL}\ra \la {\cal O}_{lR}^+\ra \label{A}
\ee
and are nonlocal in terms of both the 1DEG fermions and the local spins (hereafter I'll call them NOPs, with N for "nonlocal"). 
Since the scaling dimension of these NOPs is equal to 1/2, their vacuum expectation value $\sim \Delta^{1/2}$. To get a better understanding of NOPs, I will rewrite models (\ref{GN1},\ref{GN2}). using Abelian bosonization. For simplicity I consider the case $v_H = v_F = 1$ these modes  are equivalent to the sine Gordon model with the Lagrangian:
\bea
L = \int \rd x \Big[\frac{ (1+ J_K/\pi )}{2}(\p_{\mu}\Phi)^2 - \frac{J_K}{\pi a_0}\cos(\sqrt{8\pi}\Phi)\Big],
\eea
where $\Phi = \varphi_s +\bar\varphi_H$ for (\ref{GN1}) or $\varphi_H + \bar\varphi_s$ for (\ref{GN2}). The NOPs  (\ref{ops}) correspond to $\la \cos(\sqrt{2\pi}\Phi)\ra$. In the ground state this vacuum average may have any sign. Since only  the product (\ref{A}) enters into observable quantities, the ground state degeneracy is 2. This corresponds to the ground state degeneracy of spin S=1/2 antiferromagnetic chain.


\section{Topological phase transition}

To make sure that the ground state of model (\ref{model3}) is topologically nontrivial, I will add to the Hamiltonian competing marginally relevant interaction: 
\bea
V_{comp} = g_0({\bf F}_R{\bf F}_L + {\bf j}_R{\bf j}_L), ~~ g_0 >0. \label{comp}
\eea
This interaction is generated by  a point-like attraction in 1DEG and by the second neighbor exchange $J_{nnn} > 0.24 J$ in the Heisenberg chain. At $J_K=0$ we get two separate Hamiltonians with gapped spectrum, one for 1DEG and another for the Heisenberg chain. The ground states are topologically trivial. Now I will consider the case when both $g_0$ and $J_K$ are nonzero and demonstrate that when $g_0$ increases from zero to some critical value, the system goes through the Ising QCP. For simplicity I consider the case $v_F = v_H =v$ when the criticality is achieved at  $g_0 =J_K$. 

 Combining the interaction in (\ref{GN1},\ref{GN2}) with (\ref{comp}) and using 
the identities \cite{mybook}
\bea
j^a_R+ F^a_R = \frac{\ri}{2}\epsilon^{abc}r_br_c, ~~ j^a_R- F^a_R =\ri r_0r_a,
\eea
where $r_0, r_a ~~(a=1,2,3)$ are right-moving Majorana fermions, and similar identities for the currents of the left chirality, we rewrite Hamiltonian in the spin sector as 
\bea
{\cal H}_s = \frac{\ri v}{2}\sum_{a=0}^3(l_a\p_x l_a - r_a\p_x r_a) +\frac{1}{2}(J_K+g_0)\sum_{a>b=1}^3(r_al_a)(r_bl_b) + \frac{1}{2}(g_0 -J_K)(r_0l_0)\sum_{a=1}^3(r_al_a) \label{GN3}
\eea
Model (\ref{GN3}) is integrable \cite{ConTsv},\cite{natan}. The marginal interactions always generate mass gaps, but their mutual signs depend on the sign of $J_K - g_0$. The critical point occurs at $J_K=g_0$; at this point the singlet (the 0-th) Majorana fermion decouples from the rest and becomes massless. This is the Ising critical point. At $J_K>g_0$ the sign of the mass of the 0-th fermion is opposite to the masses of the other three and this corresponds to a topologically nontrivial  ground state. In the inhomogenious configuration when  the phases with different sign of $J_K-g_0$ touch each other, one should expect to find a zero energy Majorana mode located at the boundary. 

\section{The effective interactions}

The interchain interaction between the OPs on sites $1$ and $2$  is generated by the fusion 
\bea
U_{12} = \tilde J\int \rd^2x_2\rd^2 x_3\hat t_{12}(1)\hat t_{12}(2) {\bf N}_1(3){\bf N}_2(3)
\eea
where $\hat t_{12}$ is the tunneling operator (\ref{tunn}). On small energies the integrand becomes  
\bea
t^2\tilde J\Big\{[L^+_{1\alpha}(1){\bf N}_1(3)R^+_{1\beta}(2)]_y[R_{2\beta}(2){\bf N}_2(3)L_{2\alpha}(1)]_{y+1} - [L^+_{1\alpha}(1){\bf N}_1(3)R_{1\beta}(2)]_y[R^+_{2\beta}(2){\bf N}_2(3)L_{2\alpha}(1)]_{y+1}\Big\} \label{fusion}
\eea
where the numbers are shorthand for the coordinates $(\tau_i,x_i)$ ($i=1,2,3$). The sign difference between the two terms in (\ref{fusion}) is responsible for the appearence of $\s^z$ in (\ref{GL}). Substituting  (\ref{RL},\ref{z},\ref{G}) into (\ref{GL}) and using the fact that 
\bea
\la\la z^s_{\s}(\tau,x)[\bar z^H_{\s}]^+(0,0)\ra\ra = C\Delta^{1/2}K_0(\Delta[(\tau -\ri x/v_F)(\tau +\ri x/v_H)]^{1/2}) \equiv {\cal F}({\bf r}),
\eea
we obtain for ${\cal J}$ the following estimate:
\bea
{\cal J} \sim  \tilde J t^2\Delta^2 \int \rd^2x_2\rd^2 x_3 [{\cal F}({\bf r}_{13}){\cal F}({\bf r}_{23})]^2 \sim \tilde J(t/\Delta)^2. 
\eea
cited in the main text.


\begin{thebibliography}{99}
\bibitem{remark} In one dimension critical points are located at $T=0$ and there is only quasi long range order. Hence by order parameter operators I mean  the operators whose susceptibilities diverge at zero temperature.
\end{thebibliography}
\end{document}